\documentclass[pra,aps,twocolumn,showpacs]{revtex4}

\usepackage{graphicx}
\usepackage{bm}

\begin{document}

\title{Engineering quantum operations on traveling light beams by multiple photon addition and subtraction}

\author{Jarom\'{\i}r Fiur\'{a}\v{s}ek} 
\affiliation{Department of Optics, Palack\'{y} University, 17. listopadu 12,
77900 Olomouc, Czech Republic}

\begin{abstract}
We propose and investigate an optical scheme for probabilistic implementation of an arbitrary 
single-mode quantum operation that can be expressed as a function of photon number operator. The scheme coherently combines multiple photon addition and subtraction and is feasible with current technology. As concrete examples, we demonstrate that the device can perform approximate noiseless linear amplification of light and can emulate Kerr nonlinearity.
\end{abstract}


\maketitle

\section{Introduction}

Quantum  properties of light have attracted the attention of scientists since the early days of 
quantum mechanics. Arguably one of the main goals in the field of quantum optics is to realize highly nonlinear interactions at the few-photon level which would enable \emph{e.g.} to generate various highly  nonclassical states of light and implement advanced schemes for quantum information processing. Unfortunately, nonlinear coupling between single photons mediated by common material media is extremely weak so other approaches have to be pursued. 
One of the most promising techniques appears to be that of the measurement induced nonlinearities. 
As shown by Knill, Laflamme, and Milburn \cite{KLM01}, using passive linear optics, ancilla single photons and photon counting measurements one can emulate nonlinear coupling between single photons and implement all-optical quantum CNOT gate. During recent years, the feasibility
 of this approach has been corroborated by numerous experiments \cite{KokRMPreview}.
Moreover, these ideas have been extended also to the so-called continuous variable regime, where quantum information is encoded into states of optical modes instead of single photons. 
It was shown that linear canonical  transformations of quadrature operators  such as squeezing or quantum non-demolition  coupling can be implemented with passive linear optics, ancilla squeezed vacuum states, homodyne detection and feedforward \cite{Filip05,Yoshikawa07,Yoshikawa08}. Going beyond Gaussian operations, the basic techniques available are the addition \cite{Zavatta04} and subtraction \cite{Ourjoumtsev06} of a single photon. These operations enable to generate highly non-classical superpositions of coherent states \cite{Neergaard-Nielsen06,Wakui07,Ourjoumtsev09}, and distill and concentrate continuous-variable entanglement \cite{Opatrny00,Furusawa09}. If these operations are combined with coherent displacement,
it is possible to prepare an arbitrary single-mode state of light from initial vacuum or squeezed state \cite{Dakna99,Fiurasek05}.

As both single-photon subtraction and addition have been successfully demonstrated experimentally, it is 
interesting to investigate whether these elementary transformations can be combined to engineer some more complex useful quantum  operations on the states of traveling light beams. In this paper we propose and analyze scheme for approximate probabilistic realization of an arbitrary operation that can be expressed as a function of photon number operator $\hat{n}$.
This class of transformations includes for instance the Kerr nonlinearity described by a unitary operation  $\hat{U}=\exp(-i\phi \hat{n}^2)$, or a noiseless linear amplifier  \cite{Ralph08,Xiang09,Marek09} $\hat{Z}=g^{\hat{n}}$, where $g >1$ is the amplification gain. Our proposal is inspired by Ref. \cite{Kim08}, where a scheme that coherently combines photon subtraction and addition  has been devised for the purpose of direct verification of the bosonic commutation relations for  creation and annihilation operators.
We show that this scheme can be generalized to realize arbitrary operation $f(\hat{n})$. An essential advantage
of our approach is that we preserve the relative simplicity of the setup discussed in Ref. \cite{Kim08}. The scheme involves
only a single Mach-Zehnder interferometer and a single nonlinear crystal where parametric down-conversion occurs. 
The  precision of the approximation of a given operation $f(\hat{n})$ is controlled
by the number of photons $N$ counted by  two photo-detection blocks that form a part of the setup. In contrast, other
proposals for emulation of operations $f(\hat{n})$  require either serial \cite{Clausen03} or parallel \cite{Ralph08,Xiang09}  implementation of many basic building blocks. Our approach is thus appealing
from the experimental point of view, especially given that the core scheme has already been successfully 
realized experimentally by Zavatta, Bellini and coworkers \cite{Parigi07,Zavatta09}.

The rest of the present paper is organized as follows. In Sec. II we describe the proposed scheme and show how to determine
its parameters for a given target operation $f(\hat{n})$. In Sec. III we consider approximate realization
of a noiseless amplification. In. Sec. IV we study the emulation of Kerr nonlinearity. Finally, conclusions are drawn in Sec. V.

\section{Setup description}

The scheme for probabilistic implementation of an arbitrary transformation $f(\hat{n})$ that can be expressed as a function of photon number operator $\hat{n}$ is depicted in Fig. 1. As already mentioned in the introduction, this scheme is very similar to the setup proposed in Ref. \cite{Kim08}
for experimental testing of the bosonic commutation relation for annihilation and creation operators, $[\hat{a},\hat{a}^\dagger]=1$.
Here we consider a generalized version of that setup where $N$ photons are added to and also $N$ photons are subtracted from the input state.

\begin{figure}[!t!]
\centerline{\includegraphics[width=0.98\linewidth]{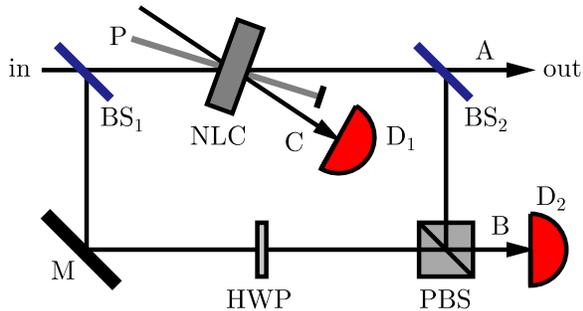}}
\caption{(Color online) Linear optical implementation of operators that are polynomials in photon number operator $\hat{n}$.
The scheme consists of unbalanced beam splitters BS$_1$ and BS$_2$, polarizing beam splitter PBS, half-wave plate HWP, mirror M,
nonlinear crystal NLC pumped by a strong laser pulse P, and two detection blocks D$_1$ and D$_2$. The detection block $D_1$  
counts the number of photons in mode $C$. Detection block $D_2$ projects the state in mode $B$ onto a specific entangled 
$N$-photon polarization state. For more details, see text.}
\end{figure}

The photon addition is achieved by feeding the state into the input signal port of a nonlinear crystal NLC 
where pairs of correlated signal and idler photons are generated in the process of noncollinear parametric down-conversion pumped by a strong coherent laser beam P \cite{Zavatta04}. The photodetector D$_1$ counts the number of emitted idler photons $N$ which is equal to the number of photons added into the signal mode. Photon counting can be achieved e.g. by employing the time-multiplexed photon number resolving detector \cite{Achilles03,Fitch03,Micuda08} where the input pulse is divided into a sequence of many pulses 
by highly unbalanced interferometers. The time-separated pulses then impinge onto avalanche photodiodes (APD) and the total number of clicks of APDs gives the number of counted photons  N. The main limitations of this detection scheme are the
finite number of time bins which does not allow for complete photon number resolution, and the non-unit detection efficiency of the APDs.
However, these limitations are not very restrictive  for the present application where the down-conversion should be operated in a regime of low
gain. The probability that $N+1$ photons are present in the output idler beam is then much lower than the probability that $N$ photons are present.
If $N$ photons are counted by D$_1$ then, with high probability,  $N$ photons were indeed 
emitted into the idler mode in NLC. In this regime, the main effect of the less-than-unit detection efficiency of APDs is a reduction of success probability of the scheme.

Mathematically, the state transformation after propagation through the crystal is described by the two-mode squeezing operator,
\begin{equation}
\hat{V}=e^{\lambda \hat{a}^\dagger \hat{c}^\dagger}(1-\lambda^2)^{\frac{1}{2}(\hat{n}_A+\hat{n}_C+1)}e^{-\lambda \hat{a} \hat{c}}. 
\label{Usqueezing}
\end{equation}
 Here $\hat{a}$ and $\hat{c}$ denote the annihilation operators of signal and idler modes, respectively, and $\lambda=\tanh s$ where
 $s$ is the squeezing constant. Assuming that the idler mode is initially in the vacuum state and that the detector D$_1$
 registers $N$ photons, the conditional transformation of the state of signal mode reads
 \begin{equation}
 \hat{A}_{N}=\frac{\lambda^N}{\sqrt{N!}} \left(1-\lambda^2\right)^{(1+\hat{n}-N)/2}\hat{a}^{\dagger N}.
 \label{AN}
 \end{equation}
In the limit of low parametric gain, $\lambda \rightarrow 0$, we obtain $\hat{A}_N \propto \hat{a}^{\dagger N}$
as desired.

The photon subtraction is performed by splitting a tiny part of the signal beam off a highly unbalanced beam splitter
followed by detection of the number of reflected photons \cite{Ourjoumtsev06,Neergaard-Nielsen06,Wakui07}. If $k$ photons are reflected from a beam splitter with amplitude transmittance $t$ and reflectance $r$ then the conditional 
transformation of the signal-mode state can be expressed as follows,
\begin{equation}
S_{k}=\frac{r^k}{\sqrt{k!}} t^{\hat{n}}\hat{a}^k.
\end{equation}
A crucial feature of the scheme in Fig. 1 is that the photon subtraction may 
occur either before or after the photon addition. The photons reflected off beam splitters BS$_1$ and BS$_2$ are
recombined on a polarizing beam splitter PBS such that their polarization states are orthogonal but all other degrees of freedom are made
indistinguishable to guarantee maximum visibility of multiphoton interference. The detector D$_2$ represents a detection block that is capable of making projection onto an $N$-photon polarization state
\begin{equation}
|\phi\rangle_{B}=\sum_{k=0}^N b_k |k\rangle_{B,H}|N-k\rangle_{B,V},
\label{phiB}
\end{equation}
where H and V denote the horizontally and vertically linearly polarized modes, respectively. Such a projection can be performed e.g. by reversing a linear optical scheme for preparation of the $N$-photon two-mode states put forward in Ref. \cite{Fiurasek02}.
The input light beam is divided into $N$ spatial modes and a polarization analysis is performed on each mode by an elementary detection block consisting of quarter- and half-wave plates, polarizing beam splitter and a pair of single photon detectors. If a single photon is detected by each detection block, and a photon in the $j$-th detection block is projected onto a polarization state
\begin{equation}
\cos\theta_j |H\rangle_B +\sin\theta_j e^{i\phi_j}|V\rangle_B,
\end{equation}
then the whole detector projects onto $N$-photon polarization state \cite{Fiurasek02,Zou04}
\begin{equation}
\prod_{j=1}^N \left(\cos\theta_j \hat{b}_H^\dagger+\sin\theta_j e^{i\phi_j} \hat{b}_V^\dagger\right) |0\rangle_{B,H}|0\rangle_{B,V}.
\label{phifactorized}
\end{equation}
Here $\hat{b}_H^\dagger$, $\hat{b}_V^\dagger$ denote creation operators of the horizontally and vertically polarized modes, respectively.
Since every $N$-photon polarization state can be expressed in the factorized form (\ref{phifactorized}), 
this detection scheme can project onto any state (\ref{phiB}). 

If $N$ photons in total are subtracted by BS$_1$ and BS$_2$ and if addition of N photons is heralded by the detector D$_1$, 
then the operation on signal mode A and the $N$-photon polarization state of auxiliary spatial mode B are given by,
\begin{equation}
\sum_{k=0}^N \hat{S}_{k}\hat{A}_{N}\hat{S}_{N-k}\otimes |k\rangle_{B,H}|N-k\rangle_{B,V}.
\label{SAS}
\end{equation}
With a slight abuse of notation, we use in Eq. (\ref{SAS}) the tensor product symbol $\otimes$ to separate operator acting on the Hilbert space of the signal mode and the $N$-photon state prepared in the auxiliary mode B.
After projecting the mode B onto the N-photon state (\ref{phiB}) we finally obtain the operation on signal mode,
\begin{eqnarray}
\hat{W}_N =C_N h^{\hat{n}}
 \sum_{k=0}^N b_k \sqrt{{N \choose k}}\left(\frac{h}{t}\right)^{k-N} \hat{a}^k \hat{a}^{\dagger N} \hat{a}^{N-k},
 \label{WNformula}
\end{eqnarray}
where $C_N=\sqrt{1-\lambda^2}\,(r\lambda)^N/N!$ and $h=t^2\sqrt{1-\lambda^2}$.
With the help of the expression 
\begin{eqnarray}
\hat{a}^k \hat{a}^{\dagger N}\hat{a}^{N-k}&= &\frac{(\hat{n}+k)!}{(\hat{n}+k-N)!}\equiv\prod_{j=k-N+1}^k (\hat{n}+j),
\label{anpolynomial}
\end{eqnarray}
we can rewrite $\hat{W}_N$ as follows,
\begin{equation}
\hat{W}_N=C_N h^{\hat{n}} P_N(\hat{n}),
\label{WNfinal}
\end{equation}
where $P_N(\hat{n})$ is a polynomial of $N$-th order in $\hat{n}$,
\begin{equation}
P_N(\hat{n})=
 \sum_{k=0}^N b_k \sqrt{{N \choose k}}\left(\frac{h}{t}\right)^{k-N} \frac{(\hat{n}+k)!}{(\hat{n}+k-N)!}.
 \label{PNpolynomial}
\end{equation}

We now prove that by properly tailoring the state (\ref{phiB}), arbitrary polynomial $P_N(\hat{n})$ can be obtained.  
We start from Eq. (\ref{WNformula}) where we normally order the creation and annihilation operators using the formula,
\begin{equation}
\hat{a}^k \hat{a}^{\dagger N}=\sum_{j=0}^k {N \choose j} \frac{k!}{(k-j)!} \hat{a}^{\dagger N-j}\hat{a}^{k-j}.
\end{equation}
In this way we obtain
\begin{equation}
P_N(\hat{n})= \sum_{j=0}^N d_j \hat{a}^{\dagger N-j} a^{N-j},
\label{PNnormal}
\end{equation}
where $d_j=\sum_{k=j}^N M_{jk} b_k$, and 
\begin{equation}
M_{jk}=  \sqrt{{N \choose k}} {N \choose j} \frac{k!}{(k-j)!} \left(\frac{h}{t}\right)^{k-N}, \quad k\geq j.
\end{equation}
By setting $M_{jk}=0$, $j>k$, we can express the relation between $d_j$ and $b_k$ in a matrix form,
\begin{equation}
\bm{d}=\bm{M} \bm{b},
\label{dMc}
\end{equation}
where $\bm{b}=(b_0,b_1,\ldots,b_N)^T$, $\bm{d}=(d_0,d_1,\ldots,d_N)^T$.
The system of equations (\ref{dMc})  for $b_k$ can be always solved. Since $M_{jj}>0$, $\forall j$ and $M_{jk}=0$, $j>k$,
we have $\det \bm{M} >0$, $\bm{M}^{-1}$ exists and we can write $\bm{b}=\bm{M}^{-1}\bm{d}$.
Any polynomial in $\hat{n}$ can be recast into the form (\ref{PNnormal}) by normally ordering the annihilation and creation operators and the coefficients $d_k$ thus unambiguously specify the polynomial. The amplitudes $b_k$ yielding the required $d_j$ can be calculated as described above which completes the engineering of the operator polynomial $P_N(\hat{n})$.

To summarize our findings so far, we have shown that with the scheme shown in Fig. 1 we can implement an arbitrary operator that can be expressed as a product of a polynomial in $\hat{n}$ and an attenuation factor $h^{\hat{n}}$ (recall that $h \leq 1$). The scheme can approximate any given operator $f(\hat{n})$
to arbitrarily high degree $N$ determined by the number of photon subtractions and additions. More specifically, the first $N+1$ terms in Taylor series expansion in $\hat{n}$ of $f(\hat{n})$ and $\hat{W}_N$ can be made equal up to  a constant prefactor. However, the success probability of the scheme will decrease approximately exponentially with increasing $N$. Practical implementations of the proposed scheme in a near future would be thus most probably limited to $N=1$ and $N=2$.

In this low-approximation regime the influence of the attenuation factor $h^{\hat{n}}$ could be particularly significant. This effective attenuation can be suppressed by employing highly unbalanced beam splitters, $t\rightarrow 1$, and working in the regime of very low parametric gain, $\lambda\rightarrow 0$. Then we get $\hat{W}_N=C_N P_N(\hat{n})$. However, in this limit the success probability of the scheme becomes very low and vanishes when $t=1$ or $\lambda=0$. 
In an experiment, optimal balance between success rate and performance of the scheme can be sought by tuning $\lambda$ and $t$.

\section{Noiseless amplifier}

As a first application of the scheme let us consider probabilistic implementation of a noiseless linear amplifier \cite{Ralph08,Xiang09},
\begin{equation}
\hat{Z}(g)=g^{\hat{n}},
\label{Zamplifier}
\end{equation}
where $g>1$ denotes an amplitude gain of the amplifier. The amplifier (\ref{Zamplifier}) acts as a non-unitary quantum filter that enhances amplitudes of the Fock states $|n\rangle$ by a factor $g^n$. Clearly, this transformation cannot be implemented exactly with a finite probability because the operator $\hat{Z}(g)$ is unbounded for any $g>1$. We can nevertheless approximately realize the transformation to $N$-th order in $\hat{n}$ by the scheme described in a previous section. In essence in this approach the amplification 
works well for low Fock states $|n\rangle$, $n<N$, but it fails for large Fock states $n \gg N$.

For the sake of simplicity, in what follows we shall assume that $h=1$, i.e. the limit of a low reflectance of beam splitters BS$_1$, BS$_2$ and low gain of the nonlinear crystal.  The influence of $h<1$ on the performance of the amplifier is that it reduces the effective gain, $g \rightarrow hg$. This can be compensated by substitution $g \rightarrow g/h$ when  determining the parameters of the scheme.

\begin{figure}[t]
\centerline{\includegraphics[width=0.8\linewidth]{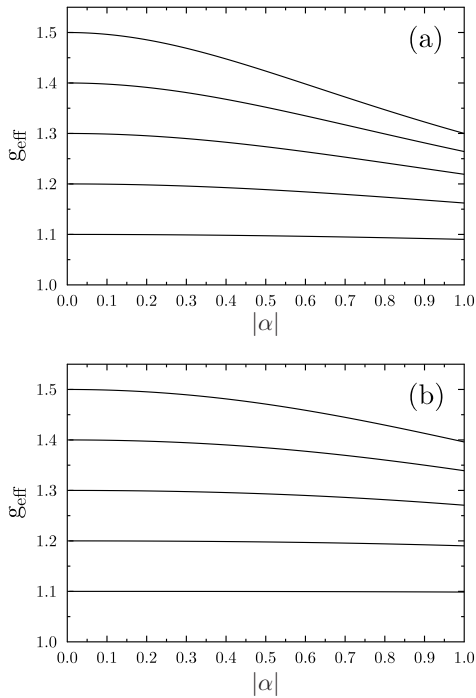}}
\caption{Dependence of the effective gain $g_{\mathrm{eff}}$ of the probabilistic amplifier on the amplitude of coherent state
$|\alpha|$ is plotted for five different nominal values of the gain, $g=1.1,1.2,1.3,1.4,1.5$ and for two levels of approximation 
$N=1$ (a) and $N=2$ (b).}
\end{figure}

\begin{figure}[!t!]
\centerline{\includegraphics[width=0.8\linewidth]{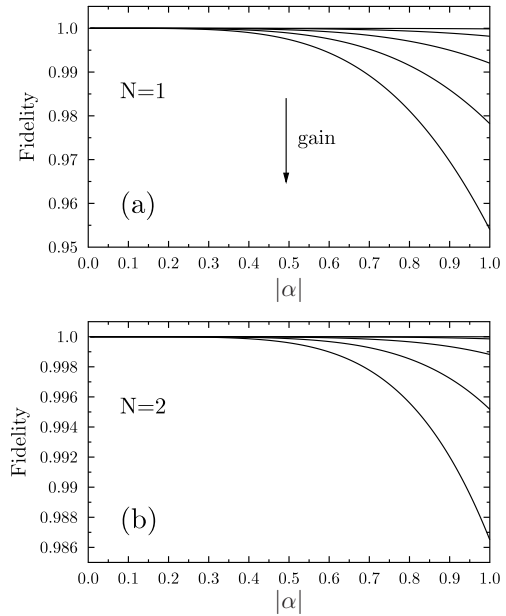}}
\caption{The fidelity of the amplified coherent states is plotted as a function of the  coherent state
amplitude $|\alpha|$ for five different gains $g=1.1,1.2,1.3,1.4,1.5$ and for $N=1$ (a) and $N=2$ (b).}
\end{figure}

We can construct a polynomial approximation to the operator $\hat{Z}(g)$ by expanding it in Taylor series in $\hat{n}$
and keeping the first $N+1$ terms. In this way we obtain a polynomial
\begin{equation}
\hat{Z}_N=\sum_{k=0}^N \frac{ d^k }{k!}\hat{n}^k,
\label{ZNapproximation}
\end{equation}
where $d=\ln g$. 
An important feature of the amplifier (\ref{Zamplifier}) is that it preserves the structure of coherent states and only amplifies their amplitude,
$\hat{Z}|\alpha\rangle \propto |g\alpha\rangle$. It turns out that the gain of the approximate amplifier (\ref{ZNapproximation})
 depends on $|\alpha|$ and we can define an effective gain of the amplifier,
\begin{equation}
g_{\mathrm{eff}}=\frac{1}{\alpha}\frac{\langle \alpha|\hat{Z}^{\dagger}_N \hat{a}\hat{Z}_N|\alpha\rangle}{\langle \alpha|\hat{Z}^{\dagger}_N\hat{Z}_N|\alpha\rangle}.
\label{geffdefinition}
\end{equation}
It is natural to require that $g_{\mathrm{eff}}=g$ in the limit $|\alpha|\rightarrow 0$. This can be achieved
if we set $d$ to be equal to a root of a polynomial equation
\begin{equation}
\sum_{k=0}^N \frac{d^k}{k!}=g.
\end{equation}
As a concrete example we now study in more detail the cases $N=1$ and $N=2$ that are experimentally feasible with current technology.
For $N=1$ we find that $d=g-1$ and the approximate polynomial reads
\begin{equation}
\hat{Z}_1=(g-1)\hat{n}+1.
\label{Z1amplifier}
\end{equation}
 For $N=2$ we must solve a quadratic equation for $d$ which yields $d=\sqrt{2g-1}-1$
and
\begin{equation}
\hat{Z}_2=(g-\sqrt{2g-1})\hat{n}^2+(\sqrt{2g-1}-1)\hat{n}+1.
\label{Z2amplifier}
\end{equation}
On inserting the operators (\ref{Z1amplifier}) and (\ref{Z2amplifier}) into Eq. (\ref{geffdefinition}) we can derive analytical formulas for the effective gain.
For $N=1$ we obtain
\begin{equation}
g_{\mathrm{eff}}=1+\frac{(g-1)[1+(g-1)|\alpha|^2]}{1+(g^2-1)|\alpha|^2+(g-1)^2|\alpha|^4}.
\end{equation}
The expression for $N=2$ is rather long and unwieldy and is not reproduced here. Instead, we plot in Fig. 2
the dependence of $g_{\mathrm{eff}}$ on $|\alpha|$ for $N=1$, $N=2$ and for five different nominal values of the gain. We can see that the effective gain decreases with increasing $|\alpha|$. Also, the performance of the scheme improves when higher-order approximation is employed, and we find that for $N=2$ the effective gain $g_{\mathrm{eff}}$ is closer to $g$ than for $N=1$. Besides the effective gain, we also calculate the fidelity of the amplified weak coherent state with the ideal target state,
\begin{equation}
F=\frac{|\langle g\alpha|\hat{Z}_N|\alpha\rangle|^2}{\langle \alpha|\hat{Z}^{\dagger}_N\hat{Z}_N|\alpha\rangle}.
\end{equation}
We plot the fidelity as a function of coherent state amplitude in Fig. 3.
For the considered range of parameters $g \leq 1.5$ and $|\alpha|\geq 1$  the fidelity exceeds $0.95$
for $N=1$ and surpasses $0.986$ for $N=2$. The scheme is thus very suitable for probabilistic error-free quantum 
cloning of weak coherent states via noiseless amplification.

\begin{figure}[!t!]
\centerline{\includegraphics[width=0.85\linewidth]{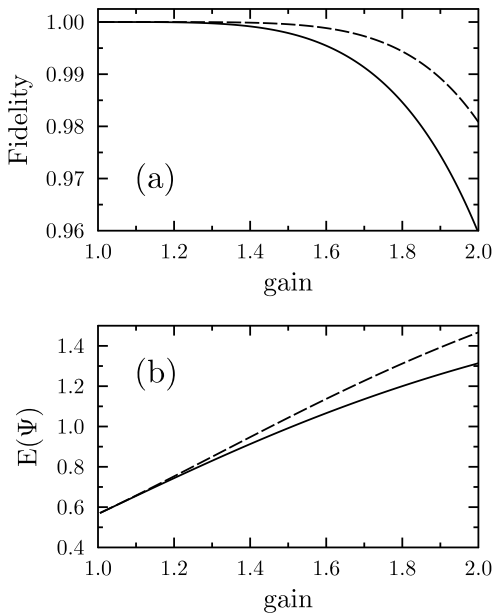}}
\caption{The figure shows dependence of fidelity (a) and entropy of entanglement (b) of the amplified 
two-mode squeezed vacuum state on the amplification gain. The parameters read $\xi=\frac{1}{3}$, 
$h\rightarrow 1$, $N=1$ (solid line), and $N=2$ (dashed line).}
\end{figure}

Another important application of a noiseless amplifier is the concentration of continuous variable
entanglement \cite{Ralph08}. Consider a two-mode squeezed vacuum state
\begin{equation}
|\Psi\rangle_{AB}=\sum_{n=0}^\infty c_n |n\rangle_A|n\rangle_B,
\label{PsiAB}
\end{equation}
where $c_n=\sqrt{1-\xi^2} \xi^n$ and $\xi$ is the  two-mode squeezing constant. 
Local noiseless amplification of one mode increases the two-mode squeezing, $\xi \rightarrow g \xi$, and thus enhances the entanglement of the state. We have investigated the effect of the approximate amplifications (\ref{Z1amplifier}) and (\ref{Z2amplifier}) on the entangled state (\ref{PsiAB}) by means of numerical calculations. In the simulations we assume the limit $h\rightarrow 1$. Generally, the attenuation factor $h^{\hat{n}}$ can be accounted for by replacing $\xi$ with rescaled squeezing constant $\xi'=h\xi$.

The results are presented in Fig.~4 that shows the dependence of fidelity 
and entanglement of the amplified state on gain $g$ for $\xi=\frac{1}{3}$ which corresponds to $3$~dB of two-mode squeezing. 
We can see that the amplification enhances the state entanglement as expected, while preserving well the structure of the
state as witnessed by high fidelity. This is further illustrated in Fig. 5 where we plot the 
Fock-state probability amplitudes  $c_n$ of a state locally amplified by operations $\hat{Z}_1$ and $\hat{Z}_2$
with $g=1.75$, as well as the amplitudes  of the initial two-mode squeezed vacuum state and the ideal amplified two-mode squeezed vacuum state with squeezing constant $g\xi$.

\begin{figure}[!t!]
\centerline{\includegraphics[width=0.95\linewidth]{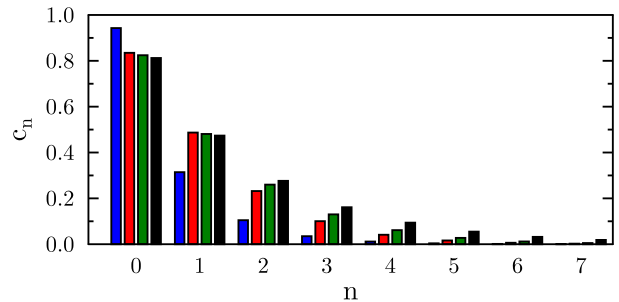}}
\caption{(Color online) The figure shows change of the Fock-state probability amplitudes $c_n$ by  noiseless amplification. For each $n$, the four bars from the left to the right 
correspond to the following states:
initial two-mode squeezed state with squeezing constant $\xi$ (blue), state amplified by operation $\hat{Z}_1$ (red), state amplified by operation $\hat{Z}_2$ (green), target two-mode squeezed state 
with squeezing constant $g\xi$ (black). The parameters are $\xi=\frac{1}{3}$ and $g=1.75$.}
\end{figure}

\section{Kerr nonlinearity}

The Kerr effect is a consequence of a nonlinear response of a medium whose index of refraction 
depends on the intensity of the light field. Quantum mechanically, this can be described by a Hamiltonian proportional to the 
square of photon number operator. The resulting unitary transformation reads,
\begin{equation}
\hat{U}=e^{i\phi \hat{n}^2},
\label{U}
\end{equation}
where $\phi$ is a dimensionless nonlinear phase shift. Kerr nonlinearity can be used to generate highly 
non-classical states of light beams and, together with Gaussian  operations, it is sufficient
for universal quantum computing \cite{Lloyd99}. With the scheme described in Sec. II we can 
implement an approximate truncated version of the transformation (\ref{U}),
\begin{equation}
\hat{U}_{2N}=\sum_{k=0}^{N}\frac{(i\phi)^k}{k!} \hat{n}^{2k}.
\label{U2N}
\end{equation}
We can generalize this expression to compensate for the attenuation factor $h^{\hat{n}}$ up to $2N$-th order by taking truncated Taylor series of operator $e^{i\phi \hat{n}^2} h^{-\hat{n}}$,
\begin{equation}
\hat{U}_{2N}(h)=\sum_{k=0}^{2N} H_k\left(\frac{-\ln h}{2\sqrt{-i\phi}}\right) \frac{(\sqrt{-i\phi}\,\hat{n})^k}{k!},
\label{U2Nhcompensation}
\end{equation}
where $H_k(y)$ denotes Hermite polynomial.
In what follows we shall assume the  limit $h\rightarrow 1$ for the sake of simplicity. Emulation of Kerr nonlinearity is more resource-demanding than
 noiseless amplification, because already the first non-trivial approximation requires subtraction and addition of two photons,
\begin{equation}
\hat{U}_2=1+i\phi \hat{n}^2.
\label{U2}
\end{equation}
The unitary matrix $\hat{U}$ is diagonal in Fock state basis, with matrix elements $U_{nn}=e^{i\phi n^2}$. The 
matrix representing the approximate operation (\ref{U2}) is also diagonal in Fock state basis but we find that
$(U_2)_{nn}=\sqrt{1+\phi^2 n^4}e^{i\Phi_n}$, where $\Phi_{n}=\arctan(\phi n^2)$. The nonlinear phase shift $\Phi_n$ does not increase quadratically with $n$ as it should  but becomes saturated, $\Phi_{n\rightarrow \infty}=\frac{\pi}{2}$. Another aspect of the approximation (\ref{U2}) is that the  Fock state amplitudes are modulated by the factors $A_n=\sqrt{1+\phi^2n^4}$ so the transformation acts also as an amplifier. These features are illustrated in Fig. 6.

\begin{figure}[!t!]
\centerline{\includegraphics[width=0.95\linewidth]{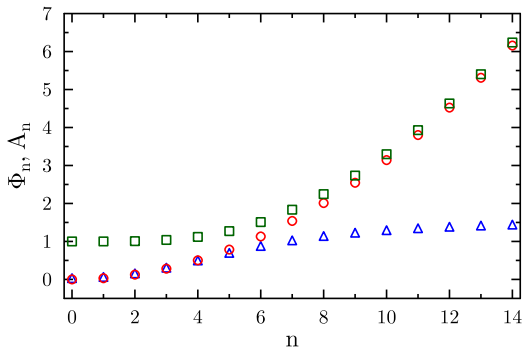}}
\caption{(Color online) Plotted are the nonlinear phase shift $\Phi_{n}$ (blue triangles) and amplitude modulation factors $A_n$ (green squares) induced by the transformation (\ref{U2}) emulating Kerr nonlinearity with $\phi=\frac{\pi}{100}$. Also shown for comparison is the phase shift $\phi n^2$ corresponding to the true Kerr nonlinearity (\ref{U}) (red circles).}
\end{figure}

We quantify the performance of the approximate operation (\ref{U2}) by the quantum process fidelity. Since 
the Hilbert space of a field mode is infinite dimensional, we shall consider fidelity $F_N$ of operation restricted to a finite dimensional subspace spanned by the first $N+1$ Fock states $|0\rangle,\ldots,|N\rangle$. We define
maximally entangled state on this subspace $|\Psi_N\rangle=\frac{1}{\sqrt{N+1}}\sum_{n=0}^N |n\rangle|n\rangle$ and 
we have
\begin{equation}
F_N=\frac{|\langle \Psi_N| (\hat{U}^\dagger\otimes \openone) (\hat{U}_2 \otimes \openone) |\Psi_N\rangle|^2}
{\langle \Psi_N| (\hat{U}_2^\dagger \otimes \openone)(\hat{U}_2 \otimes \openone)|\Psi_N\rangle},
\end{equation}
where $\openone$ denotes the identity operation.
Figure~7 shows the dependence of the fidelities $F_2$, $F_3$ and $F_4$  on the phase shift $\phi$. 
We can see that with growing $\phi$ it is increasingly more and more difficult to emulate $\hat{U}$ by $\hat{U}_2$ and the fidelities $F_N$ decrease rapidly.

\begin{figure}[!t!]
\centerline{\includegraphics[width=0.95\linewidth]{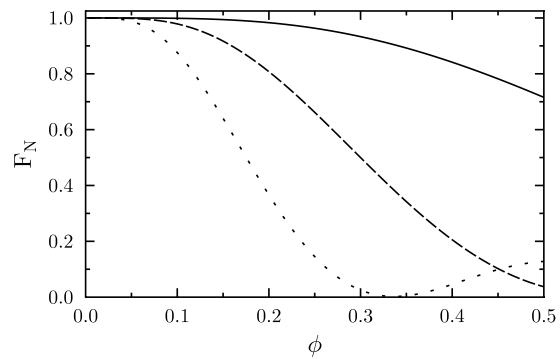}}
\caption{Dependence of the quantum process fidelity $F_N$ on the nonlinear phase shift $\phi$ 
is plotted for $N=2$ (solid line), $N=3$ (dashed line) and $N=4$ (dotted line).}
\end{figure}

This problem of fidelity decrease can be remedied if we need to emulate the Kerr nonlinearity only on a finite
dimensional subspace spanned by the first $N+1$ Fock states. More generally, any operation $f(\hat{n})$ can 
be perfectly probabilistically implemented on such subspace if we construct the polynomial approximation to this operation as follows,
\begin{equation}
 \hat{f}'_N=\sum_{k=0}^N \frac{f(k)}{h^k} \prod_{j=0,j\neq k}^N \frac{\hat{n}-j}{k-j} .
 \label{fprimeN}
\end{equation}
Note that this expression includes compensation for the attenuation factor $h$. 
In particular, on inserting $f(k)=e^{i\phi k^2}$ and $N=2$ into Eq. (\ref{fprimeN}) we obtain the following polynomial that perfectly emulates Kerr nonlinearity on a three-dimensional subspace spanned by Fock states $|0\rangle$, $|1\rangle$, and $|2\rangle$,
\begin{equation}
\hat{U}_2^\prime=\left(1-\frac{2e^{i\phi}}{h}+\frac{e^{4i\phi}}{h^2}\right)\frac{\hat{n}^2}{2} 
  -\left(3-\frac{4e^{i\phi}}{h}+\frac{e^{4i\phi}}{h^2}\right)\frac{\hat{n}}{2}+1.
\end{equation}

\section{Discussion and conclusions}

An important practical characteristics of the proposed device that we have not addressed in detail yet is its success probability. A simple order-of-magnitude estimate reads $p_{\mathrm{succ}}\approx (r\lambda \eta)^{2N}$, where $\eta$ denotes 
the detection efficiency of single-photon detectors. The quantum efficiencies of APDs exceed $50\%$. However, in the experiment, the optical beams typically need to be spatially and spectrally filtered before detection which reduces the overall detection efficiency to $10\%$ or even less.
Another important characteristics is the attenuation factor $h=t^2\sqrt{1-\lambda^2}$. We want to maximize $p_{\mathrm{succ}}$ while keeping $h$ as close to $1$ as possible. To give a concrete example, we can set 
$t^2=0.95$, $\lambda^2=0.05$, and $\eta=0.1$. For $N=1$ we obtain $p_{\mathrm{succ}} \approx 2.5\times 10^{-5}$ and $h=0.926$. Pumping the scheme with a pulsed Ti:sapphire laser with a repetition rate of $80$ MHz would then yield approximately $2000$ events per second. Second-order approximation, $N=2$, is much more challenging. For the above parameters we obtain success rate $0.05$ Hz, i.e. about $3$ events per minute which may be still acceptable for proof-of-principle experiments. Moreover, this rate can be significantly increased either by improving the detection efficiency or by using beam splitters with higher transmittance $r$. For instance, with $t^2=0.9$ and $\eta=0.2$ we obtain success rate $3$ Hz and $h=0.877$. We stress that the calculated success rates are only order-of-magnitude  estimates and would depend on the input state of the scheme as well as on details of the implemented transformation. Nevertheless, these calculations clearly confirm the experimental feasibility of the
scheme with single photon addition and subtraction and show that scheme with two photon additions and subtractions may be within the reach of present technology.

In summary, we have described and analyzed a scheme that allows to probabilistically implement arbitrary 
single-mode operation that can be expressed as a function of photon number operator. 
The method relies on coherent combination of multiple photon subtraction and addition and is experimentally feasible with present-day technology \cite{Parigi07,Zavatta09}. As concrete examples, we have shown that the scheme can work as a probabilistic noiseless amplifier and can emulate Kerr nonlinearity.  The scheme is very 
versatile and can find many applications in quantum optics and quantum information processing such as 
 implementation of quantum gates, probabilistic error-free quantum cloning, or entanglement distillation.

\acknowledgments
The author would like to thank R. Filip for stimulating discussions.
This work was supported by MSMT under projects LC06007, MSM6198959213, and 7E08028, 
by GACR under project No. GA202/08/0224, and also by the EU under the FET-Open project COMPAS (212008).

\end{document}